\documentclass[10pt,aps,pra,twocolumn,superscriptaddress,showpacs,showkeys]{revtex4-1}

\usepackage{graphicx}
\usepackage{amsmath}
\usepackage{amssymb}
\usepackage{mathtools}
\usepackage{dcolumn}
\usepackage{epsfig}
\usepackage{ifpdf}
\usepackage{color}
\usepackage{bm}
\usepackage{bbm}
\usepackage[
pdfstartview=FitH,bookmarks=true,bookmarksopen=true,breaklinks=true,colorlinks=true,linkcolor=blue,anchorcolor=blue,citecolor=blue,filecolor=blue,menucolor=blue,urlcolor=blue]{hyperref}

\begin{document}

\title{Matter wave interference of dilute Bose gases in the critical regime}
\author{Xuguang Yue}
\thanks{Present address: State Key Laboratory of Magnetic Resonance and Atomic and Molecular Physics, Wuhan Institute of Physics and Mathematics, Chinese Academy of Sciences, Wuhan 430071, China}
\affiliation{Wilczek Quantum Center, Zhejiang University of Technology, Hangzhou 310023, China}
\affiliation{Department of Applied Physics, Zhejiang University of Technology, Hangzhou 310023, China}
\author{Shujuan Liu}
\affiliation{Wilczek Quantum Center, Zhejiang University of Technology, Hangzhou 310023, China}
\affiliation{Department of Applied Physics, Zhejiang University of Technology, Hangzhou 310023, China}
\author{Biao Wu}
\affiliation{Wilczek Quantum Center, Zhejiang University of Technology, Hangzhou 310023, China}
\affiliation{International Center for Quantum Materials, School of Physics, Peking University, Beijing 100871, China}
\affiliation{Collaborative Innovation Center of Quantum Matter, Beijing 100871, China}
\author{Hongwei Xiong}
\thanks{Electronic address: hwxiong@zjut.edu.cn}
\affiliation{Wilczek Quantum Center, Zhejiang University of Technology, Hangzhou 310023, China}
\affiliation{Department of Applied Physics, Zhejiang University of Technology, Hangzhou 310023, China}

\date{\today}

\begin{abstract}
Ultra-cold atomic gases provide new chance to study the universal critical behavior of phase transition. We study theoretically the matter wave interference for ultra-cold Bose gases in the critical regime. We demonstrate that the interference in the momentum distribution can be used to extract the correlation in the Bose gas. A simple relation between the interference visibility and the correlation length is found and used to interpret the pioneering experiment about the critical behavior of dilute Bose gases [Science {\bf 315}, 1556 (2007)]. Our theory paves the way to experimentally study various types of ultra-cold atomic gases with the means of matter wave interference.
\end{abstract}

\maketitle

\section{Introduction}
\label{sec:intro}
Matter wave interference is an important branch of ultra-cold atomic gases \cite{Cronin2009RMP}. On the one hand, it demonstrates directly the coherence property of the system. For example, the spatial coherence property of Bose-Einstein condensates (BECs) in dilute gases was firstly shown experimentally with the observation of clear interference fringes for two overlapping condensates by Ketterle's group \cite{Andrews1997Sci}. On the other hand, it provides an important means to reveal the many-body physics of ultra-cold atomic gases \cite{Bloch2008RMP}. For example, it is also used to reveal the quantum phase transition from a superfluid to a Mott insulator \cite{Greiner2002Nature}, low-dimensional quantum fluctuations \cite{Polkovnikov2006PNAS, Gritsev2006NatPhys, Polkovnikov2007EPL, Hadzibabic2006Nature, Hofferberth2008NatPhys} and quantum correlations \cite{Fang:2016dm}, and quantum depletion \cite{Chang:2016fa}. Even for the Ketterle's experiment \cite{Andrews1997Sci}, its interpretation involves novel many-body physics \cite{Javanainen1996PRL, Cirac1996PRA, Castin1997PRA, Xiong2006NJP, Cederbaum2007PRL, Liu2007NJP, Masiello2007PRA, Paraoanu2008PRA}.

In the last few years, the critical \emph{phenomena} for ultra-cold atomic gases have been given intensive studies both experimentally \cite{Greiner2002Nature, Simon2011Nature, Zhang2012Sci, Hung2011Nature, Ku2012Sci} and theoretically \cite{Kato2008NatPhys, Campostrini2009PRL, Zhou2010PRL, Diehl2010PRL, Guan2011PRA, Yin2011PRA, Kuhn2012PRA, Kuhn2012PRA2, Hazzard2011PRA, Fang2011PRA}. In 2007, the matter wave interference is used to reveal the universal critical behavior of ultra-cold Bose gases in the critical regime \cite{Donner2007Sci}. Near a second-order phase transition point, the fluctuations of the order parameter dominate the behavior of the system over all length scales and get strongly correlated. This strongly correlated many-body state shows surprisingly simple and universal critical relations \cite{Zinn-Justin2002, Chaikin2000, Privman1991}. The theory of critical phenomena predicts divergent behavior of the correlation length in the critical regime. It is understandable that in the critical regime, there would be clear interference fringes if two ultra-cold atomic clouds are allowed to overlap. With this idea of matter wave interference, Ref. \cite{Donner2007Sci} gives the first experimental demonstration of the divergent and universal behavior of the correlation length for ultra-cold Bose gases in the critical regime. Recently, the Talbot-Lau interferometry \cite{Xiong2013LPL} is used to further reveal the critical regime of the ultra-cold Bose gases.

However, the relation between the measured visibility and correlation function in the off-diagonal long-range order is still an open question. A proportional relation is used in Ref.~\cite{Donner2007Sci} without proof to extract the correlation length, which plays a key role to demonstrate the divergent and universal behavior of the correlation length in the critical regime. It is the purpose of this work to study theoretically the relation between the interference visibility of two released atomic clouds from the ultra-cold Bose gas and the correlation function showing the spatial correlation for atoms at different locations. The present work has potential applications for other ultra-cold atomic gases if the interference is adopted to extract the correlation.

The manuscript is organized as follows. In Sec:~\ref{sec:momentum}, we give a brief introduction to the one-body density matrix and the correlation function, and their relation with the momentum distribution. We generally discuss the momentum distribution of two subsystems in Sec.~\ref{sec:twosubsystem}, finding the relation between the visibility and the correlation function. The result is applied to the interference experiments done by T. Esslinger in Sec.~\ref{sec:appltoexp}. We verify our simple model numerically in Sec.~\ref{sec:numerical} and conclude in Sec.~\ref{sec:summary}.

\section{Momentum distribution of dilute Bose gases in the critical regime}
\label{sec:momentum}
Let us consider the one-body density matrix of a bosonic system,
\begin{equation}
\begin{split}
  n^{(1)}(\mathbf{r}_{1},\mathbf{r}_{2})&=\left\langle\hat\Psi^{\dag}(\mathbf{r}
  _1)\hat\Psi(\mathbf{r}_2)\right\rangle\\
  &=g^{(1)}(\mathbf{r}_1,\mathbf{r}_2)
  \sqrt{n(\mathbf{r}_1)}\sqrt{n(\mathbf{r}_2)}\,,
\end{split}
\end{equation}
where $\hat{\Psi}^\dagger(\mathbf{r})$ ($\hat{\Psi}(\mathbf{r})$) is the field operator creating (annihilating) a particle at the point $\mathbf{r}$, obeying the bosonic commutation relation $[\hat{\Psi}\left(\mathbf{r}\right), \hat{\Psi}^{\dag} \left(\mathbf{r}^{\prime}\right)] =\delta(\mathbf{r} -\mathbf{r}^{\prime})$. $\langle \cdot \rangle$ considered in this work means statistical ensemble average with the quantum average for a pure state as a special case.
The density distribution is then $n(\mathbf{r})=n^{(1)}(\mathbf{r},\mathbf{r})$.
$g^{(1)}(\mathbf{r}_1,\mathbf{r}_2)$ is a dimensionless correlation function which reflects the long-range correlation for atoms at different locations.
There have been intensive theoretical studies on the expression of $g^{(1)}(\mathbf{r}_1,\mathbf{r}_2)$ for a lot of systems \cite{Zinn-Justin2002, Chaikin2000, Privman1991}. For three-dimensional uniform dilute Bose gases in the critical regime, $g^{(1)}(\mathbf{r}_1,\mathbf{r}_2)$ takes the following well-known expression \cite{Zinn-Justin2002, Chaikin2000, Privman1991}
\begin{equation}
  g^{(1)}(\mathbf{r}_1,\mathbf{r}_2)\propto\frac{\exp(-|\mathbf{r}_1-%
  \mathbf{r}_2|/\xi)}{|\mathbf{r}_1-\mathbf{r}_2|},
  \label{eq:corr}
\end{equation}
where the correlation length $\xi\propto |(T-T_c)/T_c|^{-\nu}$ with $\nu$ the critical exponent.

In the experiments of ultra-cold atomic gases, one usually measures the momentum distribution $n(\mathbf{p})=\langle \hat{\Psi}^{\dag}(\mathbf{p})\hat{\Psi}(\mathbf{p})\rangle $. Interestingly, the momentum distribution $n(\mathbf{p})$ is related directly to the one-body density matrix $n^{(1)}(\mathbf{r}_{1},\mathbf{r}_{2})$.
Therefore, the momentum distribution $n(\mathbf{p})$ is also related to the correlation function
\begin{equation}
\begin{split}
  n(\mathbf{p})=&\frac{1}{(2\pi \hbar )^{3}}\int \mathrm{d}\mathbf{r}_{1}\mathrm{d}\mathbf{r}_{2}\sqrt{n(\mathbf{r}_{1})}\sqrt{n(\mathbf{r}_{2})}\\
  &\times g^{(1)}(\mathbf{r}_{1},\mathbf{r}_{2}) e^{i(\mathbf{r}_{1}-\mathbf{r}_{2})\cdot \mathbf{p}/\hbar}.
\label{eq:norder}
\end{split}
\end{equation}
This means that the measurement of $n(\mathbf{p})$ in atomic gas experiments reveals the spatial correlation $g^{(1)}(\mathbf{r}_{1},\mathbf{r}_{2})$. We will use this general relation to consider the matter wave interference in the critical regime of  an ultra-cold Bose gas in the following sections.

\section{Momentum distribution of two subsystems}
\label{sec:twosubsystem}

\begin{figure}[ht]
  \centerline{\includegraphics[width=.45\textwidth]{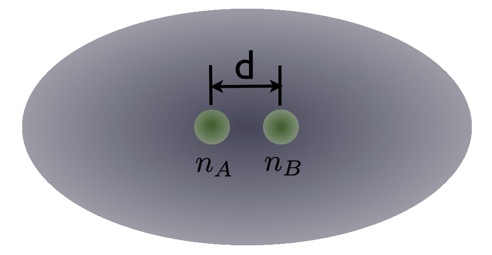}}
  \caption{(color online) Schematic representation of two atomic clouds in an ultra-cold Bose gas in the critical regime. We consider two spatially separated atomic clouds with density $n_A$ and $n_B$ separated by $\mathbf{d}$ in the interior of a Bose gas.}
\label{fig:scheme}
\end{figure}

For an ultra-cold Bose gas in the critical regime, we consider the momentum distribution of two spatially separated atomic clouds in the interior of a Bose gas, illustrated in Fig.\ref{fig:scheme}. The distance between these two atomic clouds is $d$. This is to model  the experiment in Ref.~\cite{Donner2007Sci}, where two atomic beams were outputted to extract the interference and correlation information.

We consider the case that there is no overlapping between these two atomic clouds. In this case, the overall density distribution $n(\mathbf{r})$ can be written as the sum of the density distribution of two atomic clouds, i.e.
\begin{equation}
  n(\mathbf{r})=n_A(\mathbf{r})+n_B(\mathbf{r}).
\label{eq:order}
\end{equation}
As there is no overlapping between two atomic clouds, we also have
\begin{equation}
  \sqrt{n(\mathbf{r})}\approx\sqrt{n_A(\mathbf{r})}+\sqrt{n_B(\mathbf{r})}.
\end{equation}
According to Eq. (\ref{eq:norder}), the overall momentum distribution of these two atomic clouds is
\begin{equation}
  n(\mathbf{p})=n_{AA}(\mathbf{p})+n_{BB}(
  \mathbf{p})+n_{AB}(\mathbf{p})+n_{BA}(\mathbf{p}) ,
\label{eq:np}
\end{equation}
where
\begin{equation}
\begin{split}
  n_{AA}(\mathbf{p})=&\frac{1}{(2\pi \hbar )^{3}}\int \mathrm{d}\mathbf{r}_{1}\mathrm{d}\mathbf{r}_{2}\sqrt{n_{A}(\mathbf{r}_{1})n_{A}(\mathbf{r}_{2})}\\
  &\times g^{(1)}(\mathbf{r}_{1},\mathbf{r}_{2})\exp[i(\mathbf{r}_{1}-\mathbf{r}_{2})\cdot \mathbf{p}/\hbar],\\
  n_{BB}(\mathbf{p})=&\frac{1}{(2\pi \hbar )^{3}}\int \mathrm{d}\mathbf{r}_{1}\mathrm{d}\mathbf{r}_{2}\sqrt{n_{B}(\mathbf{r}_{1})n_{B}(\mathbf{r}_{2})}\\
  &\times g^{(1)}(\mathbf{r}_{1},\mathbf{r}_{2})\exp[i(\mathbf{r}_{1}-\mathbf{r}_{2})\cdot \mathbf{p}/\hbar],\\
  n_{AB}(\mathbf{p})=&\frac{1}{(2\pi \hbar )^{3}}\int \mathrm{d}\mathbf{r}_{1}\mathrm{d}\mathbf{r}_{2}\sqrt{n_{A}(\mathbf{r}_{1})n_{B}(\mathbf{r}_{2})}\\
  &\times g^{(1)}(\mathbf{r}_{1},\mathbf{r}_{2})\exp[i(\mathbf{r}_{1}-\mathbf{r}_{2})\cdot \mathbf{p}/\hbar ],\\
  n_{BA}(\mathbf{p})=&\frac{1}{(2\pi \hbar )^{3}}\int \mathrm{d}\mathbf{r}_{1}\mathrm{d}\mathbf{r}_{2}\sqrt{n_{B}(\mathbf{r}_{1})n_{A}(\mathbf{r}_{2})}\\
  &\times g^{(1)}(\mathbf{r}_{1},\mathbf{r}_{2})\exp[i(\mathbf{r}_{1}-\mathbf{r}_{2})\cdot \mathbf{p}/\hbar].
\end{split}
\end{equation}
Here $n_{AB}(\mathbf{p})$ and $n_{BA}(\mathbf{p})$ represent the interference between two atomic clouds.

Near the critical regime, the correlation length $\xi$ is large. It is reasonable to assume the spatial size of each atomic cloud is much smaller than the correlation length $\xi$. This assumption will lead to the following two approximations.
\begin{enumerate}
\item As  the integration $\int \mathrm{d}\mathbf{r}_{1}\mathrm{d}\mathbf{r}_{2}$ in $n_{AA}(\mathbf{p})$ (or $n_{BB}(\mathbf{p})$) is over the interior of atomic cloud A (or B), $g^{(1)}(\mathbf{r}_1,\mathbf{r}_2)$ in the integral of $n_{AA}(\mathbf{p})$ (or $n_{BB}(\mathbf{p})$) can be approximated as $1$. Hence, $n_{AA}(\mathbf{p})$ and $n_{BB}(\mathbf{p})$ are the momentum distribution of each atomic cloud.
\item In the interference terms $n_{AB}(\mathbf{p})$ and $n_{BA}(\mathbf{p})$, the integrations $\int \mathrm{d}\mathbf{r}_1$ and $\int \mathrm{d}\mathbf{r}_2$ are over the interior of different atomic clouds. With this assumption, $g^{(1)}(\mathbf{r}_1,\mathbf{r}_2)$ can be approximated as a constant $g^{(1)}(d)$ with $d=|\mathbf{d}|$ the distance between two atomic clouds.
\end{enumerate}

Using these two approximations, we have
\begin{equation}
\begin{split}
  n_{AA}(\mathbf{p})\approx & \frac{1}{(2\pi \hbar )^{3}}\int \mathrm{d}\mathbf{r}_{1}\mathrm{d}\mathbf{r}_{2}\sqrt{n_{A}(\mathbf{r}_{1})n_{A}(\mathbf{r}_{2})}\\
  &\times \exp [i(\mathbf{r}_{1}-\mathbf{r}_{2})\cdot \mathbf{p}/\hbar ],\\
  n_{BB}(\mathbf{p})\approx & \frac{1}{(2\pi \hbar )^{3}}\int \mathrm{d}\mathbf{r}_{1}\mathrm{d}\mathbf{r}_{2}\sqrt{n_{B}(\mathbf{r}_{1})n_{B}(\mathbf{r}_{2})}\\
  &\times\exp [i(\mathbf{r}_{1}-\mathbf{r}_{2})\cdot \mathbf{p}/\hbar ],\\
  n_{AB}(\mathbf{p})\approx & \frac{g^{(1)}(d)}{(2\pi \hbar )^{3}}\int \mathrm{d}\mathbf{r}_{1}\mathrm{d}\mathbf{r}_{2}\sqrt{n_{A}(\mathbf{r}_{1})n_{B}(\mathbf{r}_{2})}\\
  &\times\exp [i(\mathbf{r}_{1}-\mathbf{r}_{2})\cdot \mathbf{p}/\hbar ],\\
  n_{BA}(\mathbf{p})\approx & \frac{g^{(1)}(d)}{(2\pi \hbar )^{3}}\int \mathrm{d}\mathbf{r}_{1}\mathrm{d}\mathbf{r}_{2}\sqrt{n_{B}(\mathbf{r}_{1})n_{A}(\mathbf{r}_{2})}\\
  &\times \exp[i(\mathbf{r}_{1}-\mathbf{r}_{2})\cdot \mathbf{p}/\hbar ].
\end{split}
\end{equation}
We see that the interference terms $n_{AB}(\mathbf{p})$ and $n_{BA}(\mathbf{p})$ are proportional to the correlation function $g^{(1)}(d)$.

We consider a special situation that the two atomic clouds are identical so that $n_A(\mathbf{r})=n_B(\mathbf{r}+\mathbf{d})$. In this case, we have
\begin{equation}
  n(\mathbf{p})\approx2n_{AA}(\mathbf{p})
  \left[1+g^{(1)}(d)\cos(\mathbf{p}\cdot\mathbf{d}/\hbar)\right].
\label{eq:momsimp}
\end{equation}
It is clear that the overall momentum distribution $n(\mathbf{p})$ has a term that indicates how the interference visibility depends on the dimensionless correlation function $g^{(1)}(d)$. By varying the distance between two atomic clouds, from the measurement of the density distribution $n(\mathbf{p})$, one can get $g^{(1)}(d)$.

\section{Application to the interference experiment of two outputted atomic
beams}
\label{sec:appltoexp}

\begin{figure}[t]
  \centerline{\includegraphics[width=.45\textwidth]{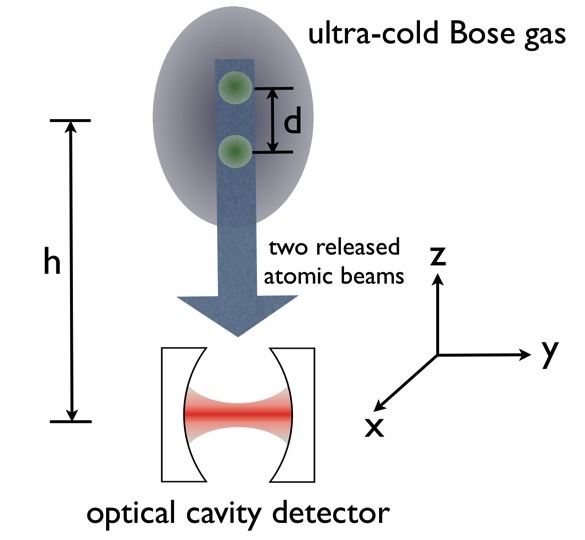}}
  \caption{(color online) Schematic representation of the interference of two released atomic beams observed by an optical cavity detector. Two clouds of atoms (green spots) with spacial separation $d$ are coupled out from an ultra-cold gas in the critical regime. Their interference pattern in time is detected with an optical cavity detector. The vertical distance between the center of the two clouds and the cavity axial line is $h$.}
\label{fig:exp}
\end{figure}

The interference is one of the most powerful means to extract the correlation or coherence property in the one-body density matrix. In Ref. \cite{Donner2007Sci}, the interference between two released atomic beams was measured to extract the spatial correlation function $g^{(1)}(\mathbf{r}_1, \mathbf{r}_2)$, as illustrated in Fig. \ref{fig:exp}. In this experiment, two atomic beams were outputted from the ultra-cold atomic gas in the critical region. The released atoms propagated downward because of gravity. Besides the downward propagation, these two atomic clouds expanded and finally overlapped with each other. The interference effect was then detected from the measured density distribution of these two overlapping atomic clouds. At the temperatures near the critical temperature, a series of visibilities $V(d)$ of the interference fringes were measured by varying the initial distance $d$ between two outputted atomic clouds.

In Ref.~\cite{Donner2007Sci}, the correlation length $\xi$ was obtained by fitting the measured visibilities with $A\exp(-d/\xi)/d$. By varying the temperature, the relation between $\xi$ and the temperature was obtained, revealing the critical exponent. However, in Ref.~\cite{Donner2007Sci}, the proportional relation between the visibility $V(d)$ and $A\exp(-d/\xi)/d$ was used as an assumption without proof. We now consider this problem with the general theory developed in previous sections.

For two outputted atomic clouds in the critical region, the density distribution of these two atomic clouds is given by Eq. (\ref{eq:order}). We consider the case that the distance between these two atomic clouds is sufficiently large so that the overlapping between $n_A(\mathbf{r})$ and $n_B(\mathbf{r})$ just before the output coupling can be omitted. This should not be confused with the final overlapping after the followed downward propagation and free expansion.

We consider the simplest case where the two outputted atomic clouds are identical so that $n_A(x,y,z)=n_B(x,y,z-d)$. We further assume that
\begin{equation}
  n_{A,B}(\mathbf{r})=n_{A,B}(z)n_{\perp}^{0}(x,y),
\end{equation}
where $n_{\perp}^{0}(x,y)$ is the density profile of the two clouds in the perpendicular direction, which can be assumed to be Gaussian, and satisfies $\int n_{\perp}^{0}(x,y)\mathrm{d}x\mathrm{d}y=1$. From Eq. (\ref{eq:momsimp}), the momentum distribution along $z$ direction is then
\begin{equation}
\begin{split}
  n(p_z)&\approx\int n(\mathbf{p})\mathrm{d}p_xd\mathrm{p}_y\\
  &=2n_z(p_z)\left[1+g^{(1)}(d)
  \cos\left(\frac{p_zd}{\hbar}\right)\right],
\end{split}
\end{equation}
where $n_z(p_z)=|\int \mathrm{d}z\sqrt{n_A(z)}\exp(ip_zz\hbar)|^2/2\pi\hbar$. The interference term in the momentum distribution is clearly given by the term
$g^{(1)}(d)\cos({p_zd}/{\hbar})$.

As illustrated in Fig.~\ref{fig:exp}, in the experiment of Ref.~\cite{Donner2007Sci}, two released atomic clouds propagate downward because of gravity. Because two released atomic clouds locate at the center of the harmonic trap, we can use uniform gas approximation to analyze the critical correlation. Assume that the distance between the ultra-cold Bose gas and the high-finesses optical cavity detecting the flux of the released atomic clouds is $h$. The velocity of the atomic clouds relative to the optical cavity is $v_0=\sqrt{2gh}$ with $g$ being the gravitational acceleration.  The free-fall time is $t_0\sqrt{2h/g}$. The measured atomic density distribution can be mapped to the momentum distribution with $z(t)\equiv p_zt/m$ for long-time free expansion. In this case, when the atomic clouds arrive at the optical cavity at time $t_0$, in the center-of-mass coordinate of two atomic clouds, we have the spatial density distribution
\begin{equation}
  n(z,t_0)=2n_z\left(p_z\equiv \frac{zm}{t_0}\right)\left[1+g^{(1)}(d)\cos\left(\frac{
  mdz}{\hbar t_0}\right)\right].
\end{equation}

In Ref.~\cite{Donner2007Sci}, they measured the flux with the high-finesse optical cavity detector. In addition, two outputted atomic beams were continuous. When the energy difference $\Delta E$ \cite{Bourdel2006PRA} per atom between two outputted atomic beams besides the gravitational potential is considered, in the laboratory frame of reference, the flux detected by the optical cavity becomes
\begin{equation}
  F(t)=F_0\left[1+g^{(1)}(d)\cos\left(\frac{mgd}{\hbar}t+\frac{\Delta E}{\hbar}t
  +\phi\right)\right].
\end{equation}
Here $F_0$ is a constant for continuous and uniform output of the atomic beams and $\phi$ is a fixed phase difference \cite{Bourdel2006PRA}.

From the above equation, we finally get the following proportional relation between the interference visibility $V$ and $g^{(1)}(d)$,
\begin{equation}
  V=g^{(1)}(d).
\end{equation}
This proportional relation does not depend on the energy difference and phase difference between two atomic beams. The period of the flux is
\begin{equation}
  T=\frac{2\pi\hbar}{mgd+\Delta E}.
\end{equation}

It is worthwhile to point out that, in the above derivation, there are two implicit assumptions: (i) In the output and the following downward propagation of the atomic beams, the momentum distribution of the atomic beams in the center-of-mass coordinate does not change. (ii) During the output of the atomic beams from the ultra-cold atomic gas, the collisions to re-establish thermal equilibrium of the whole system can be omitted. When the matter wave interference is used to reveal the spatial correlation of the system in the critical regime, the experiment should be designed to satisfy these two assumptions.


\section{Numerical simulation}
\label{sec:numerical}

In order to verify our simple while intuitive model, we solve Eq.~(\ref{eq:norder}) numerically with the experimental parameters described in Ref.~\cite{Donner2007Sci}. Specifically, we start with the initial density (\ref{eq:order}) with Gaussian distributions $n_{A,B}(\mathbf{r})=\frac{1}{\sqrt{\pi\Delta^2}}e^{-(\mathbf{r}\pm \mathbf{d}/2)^2/\Delta^2}$ separated by distance $d=|\mathbf{d}|$. Assuming the width $\Delta$ of the Gaussian distributions to be much smaller than their separation $\Delta\ll d$, we have
\begin{equation}
\begin{split}
  \sqrt{n(\mathbf{r})}\simeq & \frac{1}{\sqrt[4]{\pi\Delta^2}}\left\{\exp\left[-\frac{(\mathbf{r}+\mathbf{d}/2)^2}{2\Delta^2}\right]\right.\\
  &\qquad\qquad+\left.\exp\left[-\frac{(\mathbf{r}-\mathbf{d}/2)^2}{2\Delta^2}\right]\right\}
\end{split}
\end{equation}
Substituting it into Eq.~(\ref{eq:norder}), we can get its momentum distribution
\begin{equation}
    n(\mathbf{p}) = \frac{1}{(2\pi\hbar)^3}\int \mathrm{d}\mathbf{s} n_T(\mathbf{s})
    g^{(1)}(s)e^{i\mathbf{s}\cdot\mathbf{p}/\hbar},
\label{eq:npa}
\end{equation}
with
\begin{equation}
\begin{split}
  n_T(\mathbf{s})=&2\exp{\left(-\frac{s^2}{4\Delta^2}\right)}\\
  &+\exp{\left[-\frac{(\mathbf{s}+\mathbf{d})^2}{4\Delta^2}\right]}+\exp{\left[-\frac{(\mathbf{s}-\mathbf{d})^2}{4\Delta^2}\right]}.
  \end{split}
\end{equation}
Here $s=|\mathbf{s}|=|\mathbf{r}_1-\mathbf{r}_2|$. In the derivation of Eq.~(\ref{eq:npa}), we have carried out the integration over $\mathbf{R}=(\mathbf{r}_1+\mathbf{r}_2)/2$. As can be seen, the momentum distribution is composed of three terms, each one being the Fourier transform of a Gaussian function centered at $0, \pm d$ multiplied by the correlation function $g^{(1)}(s)$, respectively. In reminiscence of Eq.~(\ref{eq:np}), one knows that the first terms in the right hand side of the above equation is the momentum distribution of each cloud, while the second and the third term originates from the interference between the two clouds in momentum space, respectively.


As illustrated in Fig.~\ref{fig:exp}, assume that the center of the two clouds are both in the $z$-axis. Given the distribution width $\Delta$, the distance $d$ between them, the temperature $T$ of the atomic gas and the correlation length $\xi$, we finish the integration in Eq.~(\ref{eq:npa}) numerically and get the momentum distribution, from which we obtain the visibility of the interference pattern centered at $p_z=0$.

It is worth pointing out that, in calculating the integration, we have encountered problems at small $s$ since the correlation function~(\ref{eq:corr}) can be used only when $|\mathbf{r}_1-\mathbf{r}_2|\gg\lambda_T$. Here $\lambda_T=\sqrt{2\pi\hbar^2/(mk_BT)}$ is the thermal de Broglie wavelength of the atomic gas at temperature $T$ with the Boltzmann constant $k_B$ and the atomic mass $m$. We overcome this problem by using the following form of the correlation function in the critical regime when $\xi>\lambda_T$
\begin{equation}
  g^{(1)}(s) = \left\{ \begin{array}{l}
  \exp \left(-\pi s^2/\lambda_T^2\right), \quad if \quad s \le {s_c};\\[8pt]
  s_0/r\exp\left(-s/\xi\right), \quad if \quad s > {s_c},
  \end{array} \right.
\end{equation}
with $s_c=\lambda_T\left(\eta +\sqrt{\eta^2+8\pi}\right)/4\pi$ and $s_0=s_c \exp\left(s_c/\xi-\pi s_c^2/\lambda_T^2 \right)$ being two parameters determined by $\eta=\lambda_T/\xi$, the ratio of the de Broglie wavelength $\lambda_T$ and the correlation length $\xi$. $s_c$ and $s_0$ are obtained by solving the continuity equation of $g^{(1)}(s)$ and its first order derivative. In this form, $g^{(1)}(s)$ maintains a smooth function up to its first order derivative on the whole $s$-axis, besides the correct condition $g^{(1)}(0)=1$. Beside, when the temperature is far above the critical temperature, $\lambda_T\gg\xi$, and $s_c\rightarrow\infty$, the correlation function is then totally dominant by the Gaussian part. In the critical regime, $\xi\gtrsim\lambda_T$, $s_c$ is then on the order of several times of the $\lambda_T$. The long-range $e^{-s/\xi}/s$ tail then becomes important.

\begin{figure}[t]
  \centerline{\includegraphics[width=.6\textwidth]{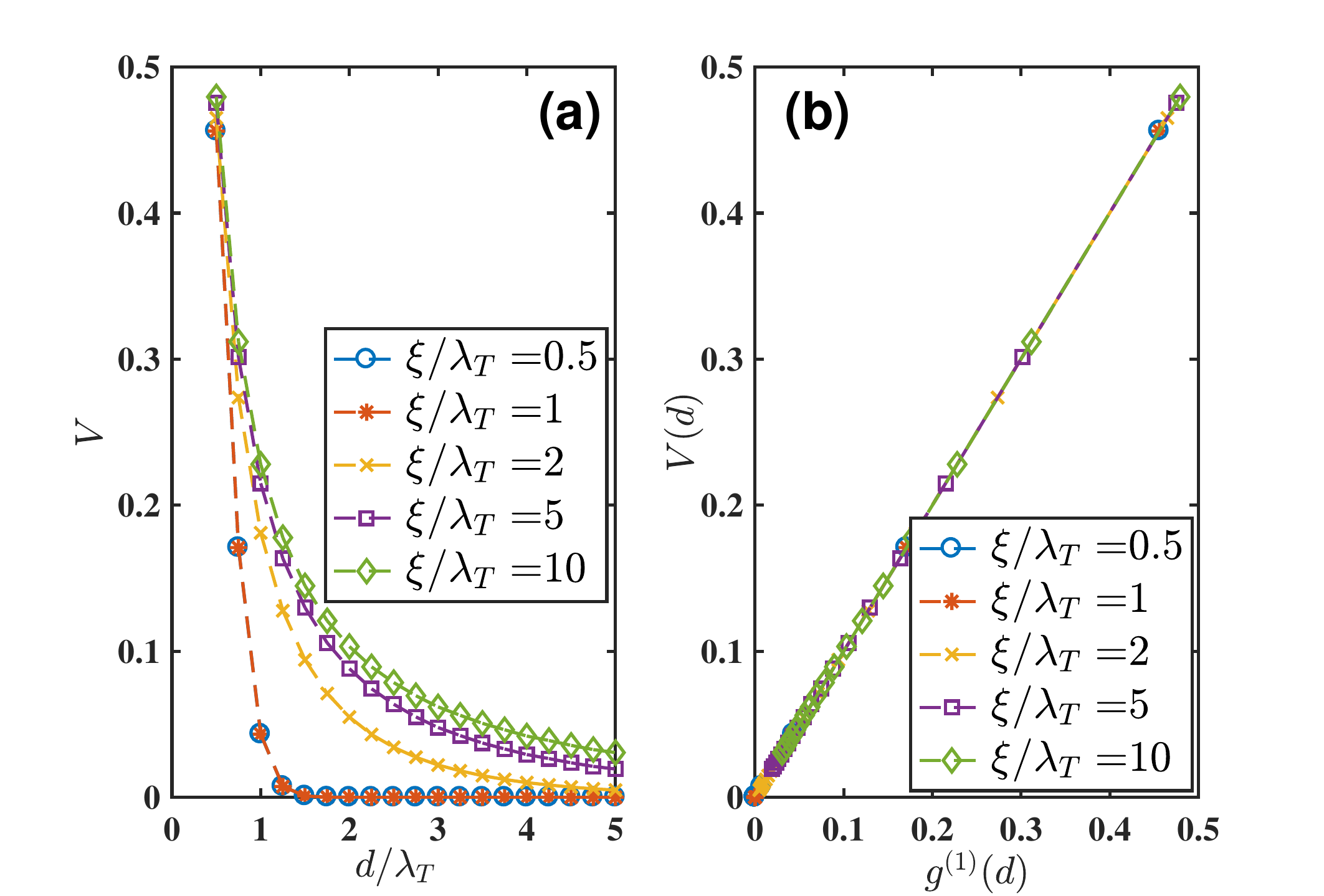}}
  \caption{(color online) (a) Dependence of the visibility $V$ on the distance $d$ between the two atomic clouds, for different correlation length $\xi$. (b) Relation of the visibility $V(d)$ with the correlation function $g^{(1)}(d)$ at different $\xi$. Here the symbols are numerical integration results and the dashed lines are given as guides to the eye.}
\label{fig:num}
\end{figure}

With the above form of the correlation function, we can perform the integration of $\mathbf{s}$ over all space including $\mathbf{s}=0$. In the previous analytical approximation in Sec:\ref{sec:twosubsystem}, however, we assume that the interior of the cloud is fully coherent. In Fig.~\ref{fig:num}(a), we show the dependence of the visibility on the distance $d$ for different correlation length $\xi$ when $\Delta\ll d$ is well satisfied. When $\xi\le\lambda_T$, the visibility drops quickly as the distance $d$ between the two clouds is increased. Besides, the visibility does not change with $\xi$, since in this case the correlation function $g^{(1)}(s)=\exp(-\pi s^{2}/\lambda_{T}^{2})$ is unrelated to $\xi$. When $\xi>\lambda_T$, a typical $e^{-r/\xi}/r$ behavior of the dependence of $V$ on $d$ is obtained. We fit the dependence with the correlation function $\xi$ being the fitting parameter as in the experiment, and find good consistency with our input value of $\xi$. We also compare the visibility with the correlation function $g^{(1)}(d)$. For different $\xi$, the visibility $V(d)$ is a linear function of $g^{(1)}(d)$. Furthermore, as can be clearly seen from Fig.~\ref{fig:num}(b), the dependence of $V(d)$ on $g^{(1)}(d)$ at different $\xi$ coincides and becomes a single line. With this result, we verify that under the conditions of $\Delta \ll d$ and $\Delta\ll\xi$, the visibility is indeed proportional to the correlation at the distance of the two clouds.

\section{Summary and discussion}
\label{sec:summary}

In summary, we have studied the momentum distribution of the ultra-cold Bose gas in the critical regime. We have found that the momentum distribution is directly relate to the correlation function of the system and its interference can be used to extract the correlation length. We have used this theory to explain the experiment in  Ref.~\cite{Donner2007Sci}. It is clear that the theory developed here can be also applied to consider the phase transition of other systems, such as the phase transition for ultra-cold Bose gases with optical lattices, low-dimensional phase transition, ultra-cold fermionic and molecular gases.

\section*{Acknowledgment}
This work is supported by the National Natural Science Foundation of China (Grant No. 11504328, 11274024, 11334001) and the National Basic Research Program of China (Grants No. 2013CB921903 and 2012CB921300)

\addcontentsline{toc}{chapter}{References}
\bibliography{ref}

\end{document}